\documentclass[final,3p,times]{elsarticle}

\usepackage{amssymb}
\usepackage{amsmath}

\usepackage{lineno}
\journal{Journal of Subatomic Particles and Cosmology}

\begin{document}

\begin{frontmatter}



\title{Anisotropic flow measurements in OO and Ne--Ne collisions with ALICE}

\author{Vojtěch Macháček on behalf of the ALICE collaboration}
\affiliation{organization={Niels Bohr Institute, University of Copenhagen},
             addressline={155 Jagtvej},
             city={Copenhagen},
             postcode={2200},
             country={Denmark}}
\begin{abstract}
The LHC results have demonstrated that small collision systems can exhibit collective behavior reminiscent of that~observed in heavy-ion collisions, opening a new frontier for exploring the connection between nuclear structure and collective dynamics at ultra-relativistic energies. However, the interpretation of collectivity in small collision systems is limited by understanding the proton subnucleonic structure. The ${}^{16}$O and ${}^{20}$Ne collisions delivered by the LHC offer an opportunity to study the influence of nuclear geometry on anisotropic flow in systems where the nuclear structure is well-constrained. In this proceeding, measurements of the anisotropic flow of charged particles, $v_n$, in OO and Ne--Ne collisions involving up to eight-particle correlations are presented, using the multi-particle cumulant method. Furthermore, the first reconstruction of the probability distribution function of elliptic flow $P(v_2)$, and the first measurement of the ratio $v_3\{4\}/v_3\{2\}$ in light-ion collision systems are reported. These measurements provide deeper insight into the role of nuclear geometry-driven flow in light-ion collisions.
\end{abstract}



\begin{keyword}


quark--gluon plasma \sep anisotropic flow \sep light-ion collisions \sep ALICE
\end{keyword}

\end{frontmatter}



\section{Introduction}
\label{Introduction}
The main goal of the ultra-relativistic heavy-ion collision program at the LHC is to study the quark--gluon plasma (QGP), a deconfined state of QCD matter belived to have existed in the early Universe \cite{ALICE-Journey}. In recent years, signatures of anisotropic flow---one of the key observables to study dynamical properties of the QGP---have also been observed in small collision systems such as pp and p--Pb collisions \cite{PRL123, NearSideRidge}. These observations open new possibilities to investigate the influence of nuclear structure on collective behavior of particles produced during the QGP expansion. Ultra-relativistic collisions of light ions such as ${}^{16}$O and ${}^{20}$Ne provide a unique probe for such studies because their nuclear structures are well understood from low-energy physics experiments \cite{NuclearStructure}. The first measurements of anisotropic flow in these systems provide evidence of nuclear geometry-driven flow in OO and Ne--Ne collisions \cite{LightIonPaper,ATLAS_LightIons}.\par
Anisotropic flow coefficients, $v_n$, arise from the Fourier decomposition of the particle azimuthal distribution with respect to the symmetry plane $\Psi_n$:
\begin{equation}
    \frac{\text{d}N}{\text{d}\varphi}\varpropto1+2\sum_{n=1}v_n\cos\left(n\left(\varphi-\Psi_n\right)\right)
\end{equation}
To unravel the underlying distribution of the anisotropic flow coefficients, it is necessary to measure multi-particle correlations sensitive to higher moments of the distribution and to event-by-event fluctuations. Measurements of multi-particle correlations in large collision systems \cite{FlowALICE} have revealed a fine splitting among results obtained using correlations involving more than four particles, suggesting the presence of non-zero higher-order moments of the~underlying probability distribution function. The elliptic power function parametrisation was proposed as a suitable function to describe the initial eccentricity in both central and peripheral nucleus-nucleus collisions \cite{Yan_2014}. This probability distribution function, $P(v_2)$, was studied in large collision systems \cite{EliptcPowerFunctionALICE,ElipticPowerFunctionATLAS, CMS_Pv2}, but it has not been previously studied in small collision systems. The results presented below complement and further extend currently available measurements of anisotropic flow in light-ion collisions.

\section{Analysis Details}
\label{AnalysisDetails}
Data recorded with the ALICE detector \cite{ALICEupgrade} during the light-ion campaign in July 2025 were analyzed. The dataset contains $\sim3\cdot10^9$ OO minimum-bias collisions and $\sim410\cdot10^6$ Ne--Ne minimum-bias collisions, both at $\sqrt{s_\mathrm{NN}}=5.36~\mathrm{TeV}$. The estimation of collision centrality is based on multiplicity measurements in the forward region ($-3.3<\eta<-2.1$) performed by one of the Fast Interaction Trigger detectors, FT0C \cite{ALICEupgrade}. Only charged particles with transverse momenta in the range $0.2<p_\mathrm{T}<3.0~\mathrm{GeV}/c$ and pseudorapidity range $|\eta|<0.8$ were used for the analysis performed by the Generic framework~\cite{GenericAlgorithm}. This framework is based on the formulation of multi-particle cumulants in terms of weighted $Q$-vectors that can be corrected for detector effects, such as non-uniform acceptance of detector or track reconstruction efficiency. The two-subevent method with pseudorapidity gap $|\Delta\eta|>1.4$ and $|\Delta\eta|>0$ was used for two-particle correlations and four-particle correlations, respectively, to suppress the short-range correlations arising from jet production and resonance decays. The standard method was applied for the rest of multi-particle correlations measurements.\par
Systematic uncertainties were estimated by varying the primary vertex position and the track selection criteria, including the distance of closest approach in the transverse plane and along the beam direction, the number of Time Projection Chamber (TPC) clusters, and the number of crossed TPC rows. The resulting relative systematic uncertainties were combined in quadrature and assigned to the final measurements. 

\section{Results}
\begin{figure}[h!]
\centering
\begin{minipage}{.5\textwidth}
  \centering
  \includegraphics[width=\textwidth]{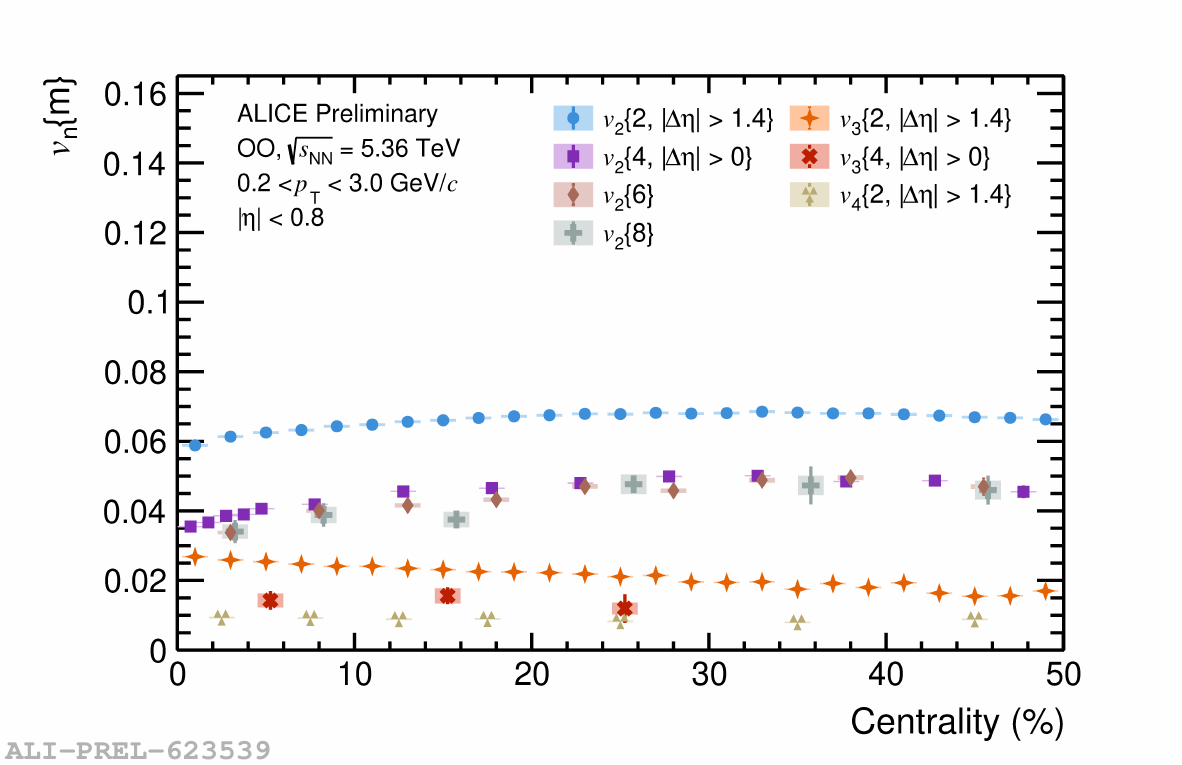}
  \label{fig:FlowOO}
\end{minipage}%
\begin{minipage}{.5\textwidth}
  \centering
  \includegraphics[width=\textwidth]{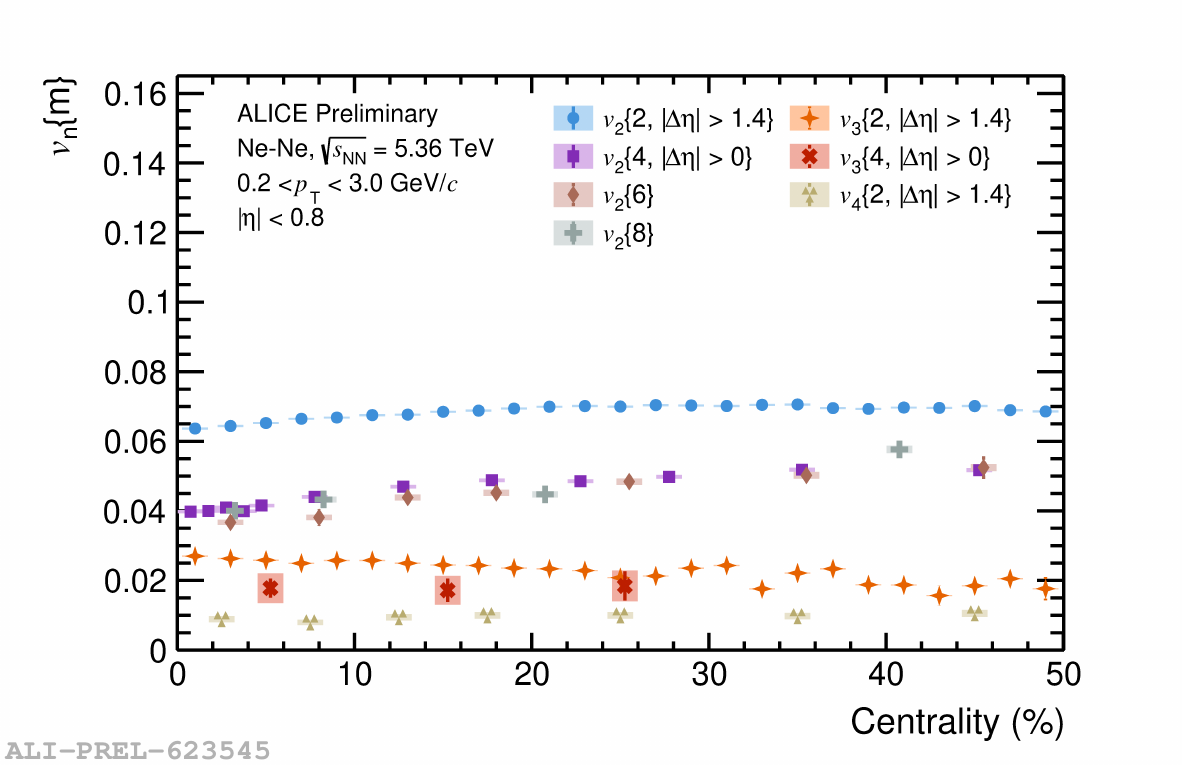}
  \label{fig:FlowNeNe}
\end{minipage}
  \caption{Anisotropic flow coefficients $v_2\{2\}$-$v_2\{8\}$, $v_3\{2\}$, $v_3\{4\}$, and $v_4\{2\}$ measured as a function of centrality in OO (left) and Ne--Ne (right) collisions.}
  \label{fig:Flow}
\end{figure}

Measurements of anisotropic flow up to the fourth harmonic and using up to eight-particle correlations are shown in Fig.~\ref{fig:Flow} for OO (left) and Ne--Ne (right) collisions. An ordering of the anisotropic flow coefficients, $v_2>v_3>v_4$, is observed in both systems. The same ordering has previously been observed in large collision systems~\cite{FlowALICE}, as well as in small collision systems in measurements performed as a function of the multiplicity of charged particles~\cite{PRL123}. The ordering of multi-particle elliptic flow coefficients $\left(v_2\{2\}>v_2\{4\}\approx v_2\{6\}\approx v_2\{8\}\right)$ is observed, pointing to an~underlying elliptic power distribution of the elliptic flow coefficient $v_2$ which has been used in large collision systems. 
\par
\begin{figure}[h!]
\centering
\begin{minipage}{.5\textwidth}
  \centering
  \includegraphics[width=\textwidth]{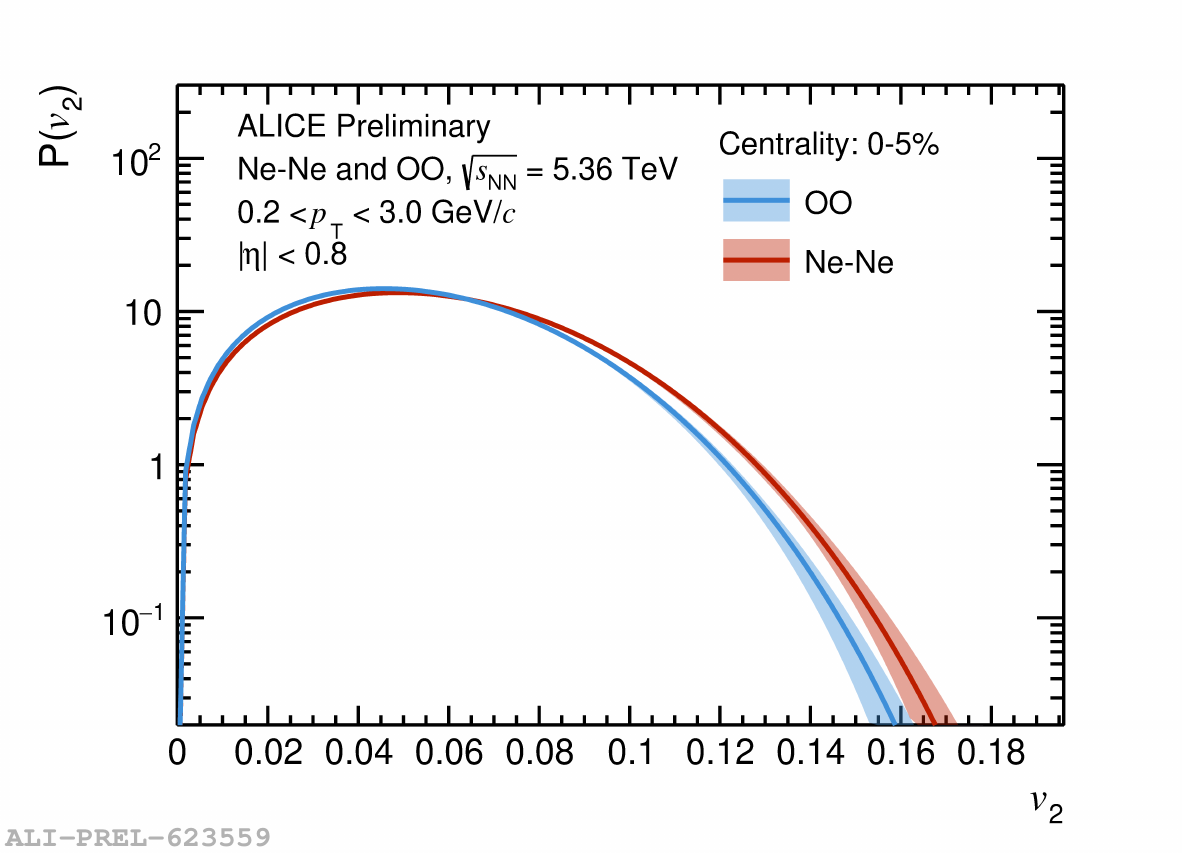}
  \label{fig:pv2}
\end{minipage}%
\begin{minipage}{.5\textwidth}
  \centering
  \includegraphics[width=\textwidth]{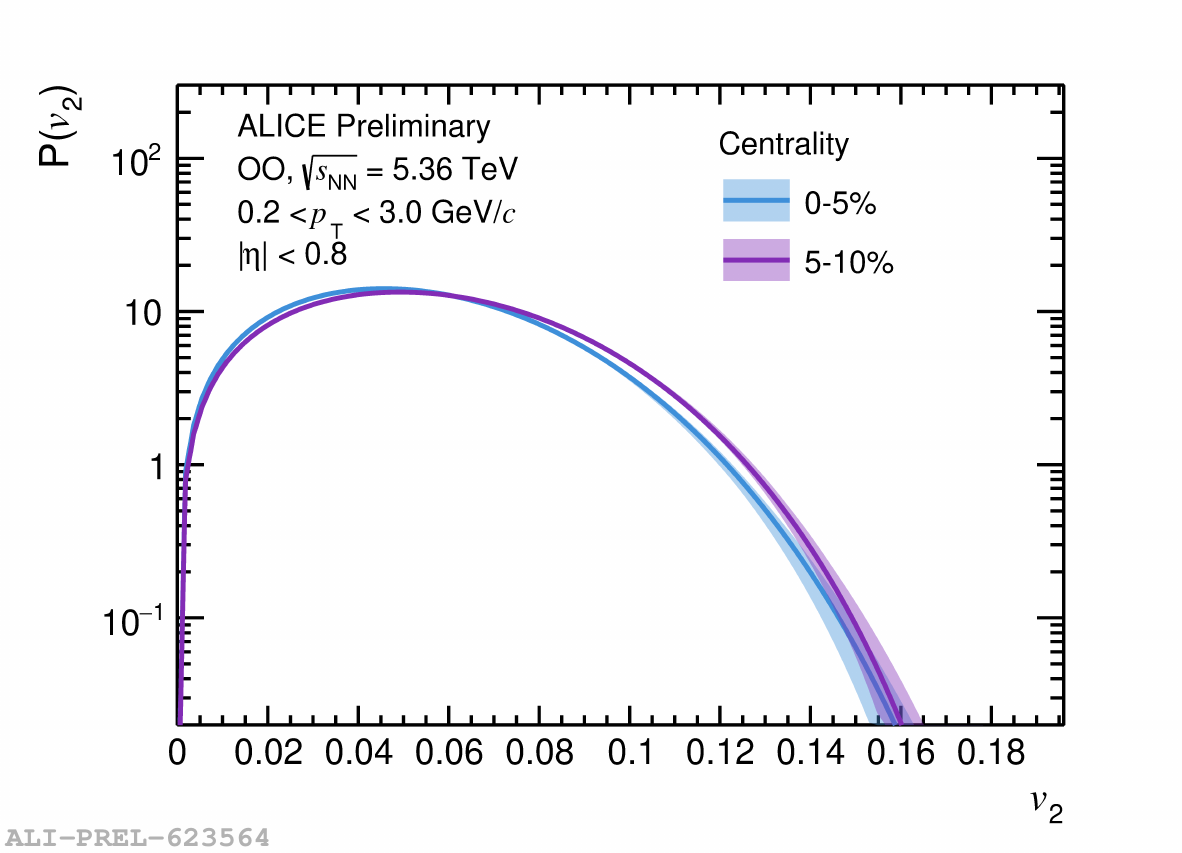}
  \label{fig:pv2_OO}
\end{minipage}
  \caption{Left: Probability distribution function of elliptic flow, $P({v_2})$, in the centrality class 0--5\% in OO and Ne--Ne collisions. Right: Evolution of the elliptic flow probability distribution function with increasing centrality in OO collisions.}
  \label{fig:Pv2}
\end{figure}

A comparison of the probability distribution function (p.d.f.) of elliptic flow, $P(v_2)$, in the most central collisions is shown in the left panel of Fig.~\ref{fig:Pv2}. In this case, the first moment of the distribution -- mean elliptic flow $\langle v_2\rangle$ -- reflects the initial eccentricity of the overlap region in the most central collisions where the colliding nuclei fully overlap. The~larger value of $\langle v_2\rangle$ observed in Ne--Ne collisions compared to OO collisions suggests a more elliptical shape of ${}^{20}$Ne in comparison to ${}^{16}$O as expected by current models of both nuclei \cite{NuclearStructure}. The evolution of the p.d.f. in OO collision in the right panel of Fig.~\ref{fig:Pv2} indicates that $\langle v_2\rangle$ increases towards semi-central collisions reflecting changes in the shape of the overlapping region.
\par
\begin{figure}[h!]
\centering
\begin{minipage}{.5\textwidth}
  \centering
  \includegraphics[width=\textwidth]{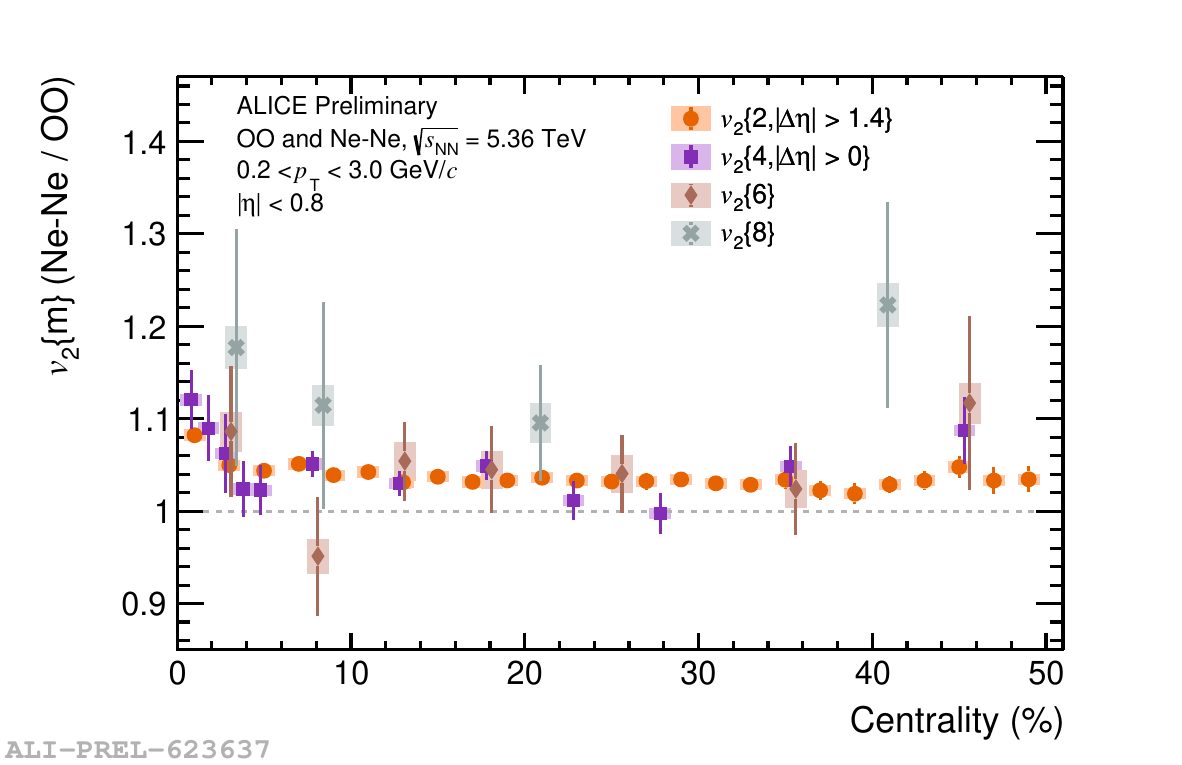}
  \label{fig:v2Ratios}
\end{minipage}%
\begin{minipage}{.5\textwidth}
  \centering
  \includegraphics[width=\textwidth]{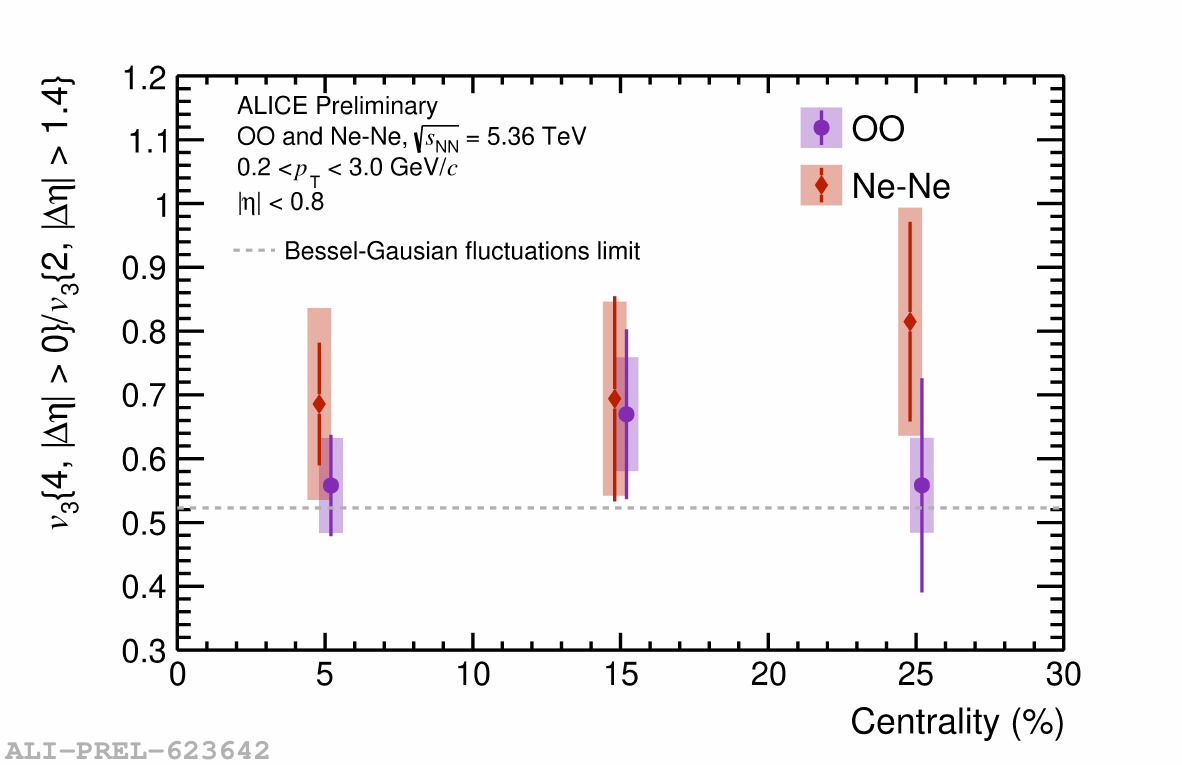}
  \label{fig:v3Ratio}
\end{minipage}
  \caption{Left: Multi-particle elliptic flow ratio $v_2\{m\}\left(\text{Ne--Ne}/\mathrm{OO}\right)$ as a function of centrality. Right: Ratio $v_3\{4\}/v_3\{2\}$ as a function of centrality in OO and Ne--Ne collisions. The dashed line indicates the universal limit expected for Bessel-Gaussian fluctuations \cite{BesselGaussFluctLim}.}
  \label{fig:Ratio}
\end{figure}

To make a comparison based purely on initial stage conditions, the ratio of multi-particle elliptic flow measurements in OO and Ne--Ne collisions is shown in the left panel of Fig.~\ref{fig:Ratio}. The deviation of the two-particle elliptic flow ratio from unity reflects differences in the nuclear shapes of ${}^{16}$O and ${}^{20}$Ne. The difference is more pronounced in the most central collisions, where the colliding nuclei fully overlap. Multi-particle elliptic flow coefficients involving more than four particles suffer from large statistical uncertainties; therefore, it is not clear whether they follow the~same trend as the two-- and four--particle elliptic flow coefficients.\par
Full reconstruction of the $P(v_3)$ distribution is not possible due to the lack of multi-particle measurements of triangular flow, $v_3$. Instead, triangular flow-fluctuations are studied in the right panel of Fig.~\ref{fig:Ratio} using the ratio $v_3\{4\}/v_3\{2\}$. In Ne-Ne collisions, the value of the ratio is enhanced relative to the universal limit of Bessel-Gaussian fluctuations. This limit is expected in central Pb--Pb collisions in which the overlap region is nearly circular. Therefore, the mean eccentricity of the collision is close to zero, and the initial elliptic power function is reduced to the Bessel-Gaussian function \cite{Yan_2014}. In that case, the anisotropic flow is predominantly driven by the fluctuations. The deviation of the ratio values from the universal limit of Bessel-Gaussian fluctuations in Ne--Ne collisions suggests the influence of the initial shape on the triangular flow.

\section{Summary}
The extensive measurements of anisotropic flow in the OO and Ne--Ne collisions using the multi-particle cumulant method have been presented. The impact of nuclear geometry on the elliptic flow $v_2$ is further studied with the~ratio $v_2\{m\}\left(\text{Ne--Ne/OO}\right)$ and the first reconstruction of the elliptic flow probability distribution function in light-ion collisions. Both measurements show an enhancement of the elliptic flow in the central Ne--Ne collisions compared to OO driven by different nuclear shapes of colliding nuclei. Signatures of the nuclear shape are also observed in the~enhancement of the $v_3\{4\}/v_3\{2\}$ ratio relative to the universal limit of Bessel-Gaussian fluctuations expected in the~central Pb--Pb collisions. Presented results provide important input for the Bayesian analysis entangling the properties of  QGP in the light-ion collisions.
\bibliographystyle{elsarticle-num}
\bibliography{ref.bib}



\end{document}